\documentclass{ws-ijmpd}
\usepackage[super,compress]{cite}
\usepackage{url}
\usepackage[dvipsnames]{xcolor}
\usepackage[utf8]{inputenc}

\usepackage{hyperref}

\begin{document}

\markboth{Isaque P. de Freitas, Gustavo O. Heymans and Nami F. Svaiter}
{Analog model for Euclidean wormholes: \\
Bose-Einstein condensate with dirty surfaces}

%
\catchline{}{}{}{}{}
%

\title{Analog model for Euclidean wormholes: \\
Bose-Einstein condensate with dirty surfaces}

\author{Isaque P. de Freitas$^1$ and Nami F. Svaiter$^2$}

\address{Centro Brasileiro de Pesquisas F\'{\i}sicas,
Rua Xavier Sigaud 150, Rio de Janeiro, 22290-180 Rio de Janeiro, RJ, Brazil.\\
$^1$isaquepfreitas@cbpf.br\\
$^2$nfuxsvai@cbpf.br}

\author{Gustavo O. Heymans\footnote{On leave from Centro Brasileiro de Pesquisas Físicas - CBPF}}

\address{Institute of Cosmology, Department of Physics and Astronomy,\\
Tufts University, Medford, Massachusetts 02155, USA
\\
olegario@cbpf.br}

\maketitle

\begin{history}
\received{Day Month Year}
\revised{Day Month Year}
\end{history}

\begin{abstract}
We study a Bose-Einstein condensate under the effects of the non-condensate atomic cloud. We model the resulting linear interaction of the condensate with the atomic gas as a quenched disorder. Using the distributional zeta function method, we obtain a representation for the quenched free energy as a series of integral moments of the partition function. Assuming that the Bose-Einstein condensate is confined between two planar surfaces, we show that random surface fields generate non-local terms in the effective action. The non-local effects in this condensed matter system define an analog model of a Euclidean wormhole. The leading contribution of the non-local interactions to the Casimir pressure is obtained.
\end{abstract}

\keywords{Bose-Einstein condensate; disorder fields;  analog model; wormholes.}

\ccode{PACS numbers:}


\section{Introduction}

The Bose-Einstein condensation is a state of matter that occurs in a gas of bosons in thermal equilibrium when the system is cooled below a certain critical temperature \cite{Bose, Einstein, AGriffinSnoke, Pethick2002, LevPitaevskii}. This condensation is a remarkable manifestation of quantum statistics and macroscopic quantum coherence. In the case of an ideal gas, the condensation is characterized by the condition that the average number of particles in the lowest single-particle level becomes a finite fraction of the total number of particles. Although the concept of Bose-Einstein condensation originally emerged as an idealization for systems governed by Bose-Einstein statistics, an analogy between this phenomenon and the behavior of liquid He$^4$ was first recognized in 1938 \cite{FLondon}. In that work, F. London suggested that the transition of liquid He$^4$ from normal fluid behavior to a superfluid state could be understood as the analog, for a liquid, of the transition that occurs in a Bose-Einstein gas at low temperatures. A more general mathematical definition of Bose-Einstein condensation, applicable not only to ideal gases but also to systems of interacting particles—where the condensation is present whenever a finite fraction of the particles occupies a single-particle quantum state—is presented in Ref. \refcite{OPenrose}. In that context, the Bose-Einstein condensate is said to be present whenever the largest eigenvalue of the one-particle reduced density matrix is an extensive, rather than an intensive, quantity. Theoretical works employing a quantum-statistical approach for liquid He$^4$ suggest how a system of interacting particles can exhibit a transition corresponding to the ideal gas transition \cite{TMatsubara, RPFeynman}. 

A Bose-Einstein condensate was produced, for the first time, in a vapor of Rb$^{87}$ atoms that was confined by magnetic fields and evaporatively cooled near a temperature of $170 \, \mathrm{nK}$ \cite{Anderson:1995gf}. In Ref. \refcite{Greytak}, a remarkable work is presented on the Bose-Einstein condensation produced from dilute atomic vapors in ultracold atoms. While ultracold quantum gases have provided condensates close to the idealization of Bose and Einstein for bosonic systems, the condensation of polaritons and magnons has introduced novel concepts of non-equilibrium condensation \cite{Hakala_2018}. Since the experimental realization of the Bose-Einstein condensate \cite{Anderson:1995gf}, several theoretical and phenomenological works have been developed with the aim of describing a trapped condensate under realistic experimental conditions. A practical problem is to be able to describe Bose-Einstein condensation in a non-ideal scenario, for example, considering a gas cloud of non-condensate atoms \cite{zaremba1999, Zaremba}. After the gas is cooled and subjected to the confinement potential $V_{\mathrm{trap}}$, approximately $10\%$ of the atomic gas remains non-condensed, acting as a cloud of atomic gas that spreads over a much larger volume compared with the Bose-Einstein condensate, which is localized at the center of the trapping potential. The aim of this paper is to discuss the effects of the non-condensate atomic gas in generating analog models of Euclidean wormholes in condensed matter \cite{Zurek:1996sj}.

Considering the absence of cosmological experiments on Hawking radiation, the Bose-Einstein condensate has been used to explore the Hawking effect by studying the propagation of excitations in the condensate \cite{Bogolyubov:1947zz, Geier:2024gvg}. Unruh has shown that the propagation of sound waves in a hypersonic fluid is equivalent to the propagation of scalar waves in black hole spacetime \cite{unruh1981}. This opens the way to the investigation of the so-called analog models on theoretical and experimental grounds \cite{NovelloABH, UnruhRalf, unruhsonic, Visser_1998, jacobson, Barcelo:2000tg, Balbinot, cadoni05, Schmitz, Garay, Jain_2007}. The analogy between wave equations in curved spacetimes and fluid systems can lead us beyond the semiclassical approximation. Based on the results obtained by Ford \cite{Ford1}, Hu and Shiokawa \cite{Hu}, and others \cite{Ford1.1, Ford2, Yu0, Thompson, Thompson2, Yu}, an analog model for quantum gravity effects was proposed in Ref. \refcite{KreinAMQGE}. The key idea of the analog model for light-cone fluctuations is to construct, in a fluid, acoustic perturbations where the sound cone fluctuates \cite{KreinAMQGE}. These authors also discussed quantum field theory with phonons in this scenario. Further discussions about analog models for light-cone fluctuations can be found in Refs. \refcite{AriasSQFTdisor,Arias:2011yg,FORD201380,Arias:2013analog}.

Based on the principles of causality and the positivity of energy, the relativistic theory of fields incorporates the principles of quantum mechanics into the classical theory of fields. In order to extend these principles to gravity, the literature has discussed modifications to standard physics at short distances. Nevertheless, it has become increasingly clear that these modifications are not sufficient to address many problems of quantum gravity. Long-range physics is argued to play a central role \cite{Giddings:2022jda}. Here, instead of discussing analog models of the behavior of a quantum field in a pseudo-Riemannian manifold, we assume the Euclidean version of quantum field theory \cite{Symanzik:1964zz, FGuerra, Glimm, JAFFE198531}. In a Euclidean theory of quantum gravity \cite{Hawking:1978jn, IshamCJ, HawkingWS, ClausKiefer, DEWITT_2008}, it is expected that the topology of spacetime can vary \cite{Anderson:1986ww}. In this context, wormholes connect two asymptotically Euclidean regions, or two parts of the same asymptotically Euclidean region \cite{COLEMAN1988643, PRESKILL1989141, Giddings:1989bq}. In this scenario, the total partition function of the system is constructed by taking into account the actions of the gravitational field and matter fields. In a theory with a single scalar field, the effects of wormholes and topology fluctuations are contained in the non-local matter field contribution.

\begin{equation}\label{eq.1}
    Z = \int [\mathrm{d} g] [\mathrm{d}\varphi] \exp\left(-S(\varphi, g) + \frac{1}{2}\sum_{i,j} \int \int \mathrm{d} \mu(x) \mathrm{d} \mu(y) \, \,  \varphi_i(x) C_{ij}(x,y)\varphi_j(y)\right).
\end{equation}

\noindent in which $[\mathrm{d} g]$ and $[\mathrm{d}\varphi]$ are functional measures, and $S(\varphi, g)$ describes the action of the gravitational and matter fields. The Riemannian $d$-volume $\mu$ is defined as $\mathrm{d}\mu = \sqrt{g}\mathrm{d} x_1...\mathrm{d} x_d$. It is expected that $C_{ij}(x,y)$ encodes the non-locality in the Riemannian manifold. In the Euclidean path integral of quantum gravity in Eq. \ref{eq.1}, we consider the effects of wormholes connecting distant regions of a single large universe. The two points connected by the wormhole are $x$ and $y$, and the effect of a wormhole is inserted through the expression $\varphi_i(x) C_{ij}(x,y)\varphi_j(y)$. It is important to distinguish such processes from ordinary propagators connecting $x$ and $y$. Ordinary processes propagate through a large region of space-time and depend on the space-time separation between $x$ and $y$. In contrast, wormholes do not depend on the space-time distance between the events at $x$ and $y$, so the coefficients $C_{ij}$ do not depend on the distance between $x$ and $y$, assuming they are distinct points in the Riemannian manifold \cite{Kleb}. This lack of dependence on space-time separation makes the wormhole amplitudes significantly different from the amplitudes for ordinary processes. In Eq. (\ref{eq.1}), we integrate over all compact topologies of space-time and focus on geometries consisting of some number of large universes connected by many wormholes.

In the present work, we propose an application of the scenario discussed in Ref. \refcite{Heymans:2023tgi} to a system of Bose-Einstein condensate in the presence of a non-condensed gas cloud.  
In the Euclidean quantum gravity scenario, one needs to average the Gibbs free energy, or the generating functional of the connected correlation functions of the system \cite{EngelhardtFERW, okuyama2021quenched}. Working with the Gross-Pitaevskii functional for the Bose-Einstein condensate, we propose that a resulting linear term arising from the interaction of the condensate with the atomic gas can be interpreted as a quenched disorder. Using the distributional zeta function method \cite{Svaiter:2016lha, Svaiter:2016jlm, Diaz:2016mto, Acosta:96.065012, Acosta-Diasorder, Heymansrestoration, Heymans2024xtk}, we obtain an effective action for the condensate with non-local terms.

The presentation of this work is organized as follows: In Sec. \ref{sec:BEC}, we construct the action functional for the Bose-Einstein condensate with contributions from the non-condensed cloud. Section \ref{sec:DZF} presents the distributional zeta function method. Multiplicative disorder is discussed in Sec. \ref{sec:randommass}. Additive disorder and the construction of the analog model are given in Sec. \ref{sec:randomfield}. Section \ref{Sec:Cas} presents the Casimir pressure due to the non-local interactions. We summarize the results and present our conclusions in Sec. \ref{sec:conc}. Throughout this work, we use $\hslash = c = k_B = 1$.

 \section{Action functional for the Bose-Einstein condensate with a non-condensed cloud}\label{sec:BEC}

The main objective of this section is to construct the action functional of the Bose-Einstein condensate, taking into account its interaction with the surrounding non-condensed cloud. The entire system is confined within a limited spatial region due to the influence of a trapping potential $V_{\mathrm{trap}}$. We achieve a unified description of the Bose-Einstein condensate and the non-condensed atomic gas system through the state of the complete system, $|\psi\rangle$, such that:
\begin{equation}\label{eq.2}
    |\psi\rangle = |\varphi \rangle + |\chi\rangle,
\end{equation}

\noindent where $|\varphi \rangle$ and $|\chi\rangle$ represent the states associated with the Bose-Einstein condensate and the non-condensed gas cloud, respectively. From these mathematical elements, we construct the order parameters corresponding to the two components of the total system. The ket $|\chi\rangle$ of the non-condensed gas is considered the thermal component of the total system, since the gas temperature is slightly higher than that of the condensate. We assume that atomic interactions occur via a Hartree-Fock pseudo-potential: $V_{\mathrm{int}} (|\mathbf{r}-\mathbf{r'}|) = g \delta(\mathbf{r}-\mathbf{r'})$, where $g$ is related to the scattering parameter, and the delta function indicates that the interaction is limited to contact between particles at nearby positions. The interaction between the gas and the condensate occurs only at the surface of the condensate. 

The Hamiltonian operator of the total system is defined as
\begin{equation}\label{eq.3}
    H_{\textrm{total}} = H_\varphi + H_{\chi} + H_{\mathrm{int}},
\end{equation}

\noindent where $H_\varphi$ and $H_\chi$ are defined respectively by 
\begin{eqnarray}\label{eq.4}
    H_\varphi = \frac{1}{2m} \sum_{l=1}^{N_\varphi} p^2(\varphi_l),
\end{eqnarray}

\begin{eqnarray}\label{eq.5}
    H_{\chi} = \frac{1}{2m} \sum_{k=1}^{N_{\chi}} p^2(\chi_k),
\end{eqnarray}

\noindent where $m$ is the mass of the atoms in the system, and $p(\varphi_l)$ and $p(\chi_k)$ are, respectively, the linear momenta of the $l$-th particle in the Bose-Einstein condensate and the $k$-th particle in the non-condensed gas. $N_\varphi$ is the number of particles in the condensate, and $N_{\chi}$ is the number of particles in the non-condensed gas. The interaction Hamiltonian, which describes interactions among all the particles in the system, is given by
\begin{equation}\label{eq.6}
    H_{\mathrm{int}} = V_{\mathrm{trap}}(\mathbf{r}) + \sum_{\substack{s,n=1 \\ n\neq s}}^{N} V_{\mathrm{int}}(|\mathbf{r}_s - \mathbf{r}_n|) = V_{\mathrm{trap}}(\mathbf{r}) + \frac{g}{2} \sum_{\substack{s,n=1 \\ n\neq s}}^{N} \delta(\mathbf{r}_s - \mathbf{r}_n),
\end{equation}

\noindent where $N$ denotes the total number of particles in the system, that is, $N = N_{\varphi} + N_{\chi}$.

The states $|\varphi \rangle$ are the direct sum of the individual states of the bosons, while the states $|\chi \rangle$ are the direct sum of the individual states of the gas particles,
\begin{equation}\label{eq7}
     |\varphi\rangle = \bigoplus_{l=1}^{N_\varphi} |\varphi_l \rangle, \quad |\chi\rangle = \bigoplus_{k=1}^{N_{\chi}} |\chi_k \rangle, \quad \mathrm{and} \quad |\psi\rangle = \bigoplus_{q=1}^N |\psi_q \rangle.
\end{equation}

\noindent In the mean-field approximation, $|\psi_q\rangle = |\psi_{q'}\rangle$ for any $q$ and $q'$. This approximation is valid for low-density systems. Consequently, the only interaction between particles is a contact interaction, described by the Hartree-Fock potential.

As usual, we decompose the field $\langle \mathbf{r}, t|\psi \rangle = \psi(\mathbf{r},t)$ into a classical field $\psi(\mathbf{r})$ and a quantum fluctuating part, represented by $\omega(\mathbf{r},t)$, as follows:
\begin{equation}
\psi(\mathbf{r},t) = e^{i\mu t}\left[\psi(\mathbf{r}) + \omega(\mathbf{r},t)\right]
\end{equation}

\noindent where $\mu$ is the chemical potential. For the stationary solution, we adopt the classical field approximation by disregarding $\omega(\mathbf{r},t)$, since $|\psi(\mathbf{r})| \gg |\omega(\mathbf{r}, t)|$. In this approximation, the expected value of the total Hamiltonian, $\mathcal{H}_{\textrm{total}}$, is given by
\begin{eqnarray}\label{eq8}
    \mathcal{H}_{\textrm{total}}  &=& \int \mathrm{d} \mathbf{r} \left[N_\varphi \varphi^*(\mathbf{r}) \left(-\frac{\Delta}{2m} + V_{\mathrm{trap}}(\mathbf{r}) \right)\varphi(\mathbf{r}) \right.  \\ 
    && + \left. N_{\chi}\chi^* (\mathbf{r})\left(-\frac{\Delta}{2m} + V_{\mathrm{trap}}(\mathbf{r})\right)\chi(\mathbf{r})\right]  \nonumber +\mathcal{H}_2.
\end{eqnarray}
In the above equation, $\varphi(\mathbf{r})$ and $\chi(\mathbf{r})$ represent the degrees of freedom of the condensate and the atomic cloud, respectively. The contribution of the interaction term $\mathcal{H}_2$ between the condensate and the gas cloud to the total Hamiltonian is given by 
\begin{eqnarray}\label{eq9}
    \mathcal{H}_{2} &=& \frac{g}{2}N(N-1) \int {\rm d} \mathbf{r} \Big[|\varphi(\mathbf{r})|^4 +  2\chi^*(\mathbf{r}) \varphi(\mathbf{r})|\varphi(\mathbf{r})|^2 + 2\chi(\mathbf{r})\varphi^*(\mathbf{r})|\varphi(\mathbf{r})|^2  \nonumber\\
   & & +  \varphi(\mathbf{r})  \chi^{*}(\mathbf{r})  \varphi^{*}(\mathbf{r}) \chi(\mathbf{r}) + 4 |\varphi(\mathbf{r})|^2|\chi(\mathbf{r})|^2 + 2\varphi(\mathbf{r}) \chi^*(\mathbf{r})|\chi(\mathbf{r})|^2  \nonumber\\
    & &+2\varphi^*(\mathbf{r}) \chi(\mathbf{r})|\chi(\mathbf{r})|^2 + |\chi(\mathbf{r})|^4\Big].
\end{eqnarray}
For $N \gg 1$, $\mathcal{H}_{2}$ can be rewritten as
\begin{align}\label{eq10}
    &\mathcal{H}_{2} = \frac{9}{10}N \int \mathrm{d} \mathbf{r} \Big[\varphi^*(\mathbf{r})\, \eta(\mathbf{r})  \varphi(\mathbf{r}) + g_\varphi |\varphi(\mathbf{r})|^4 + \varphi(\mathbf{r}) h_{\chi}^*(\mathbf{r}) + \varphi^*(\mathbf{r}) h_{\chi}(\mathbf{r})  \nonumber \\
     &+  \varphi^2(\mathbf{r}) m^*(\mathbf{r})+ {\varphi^*}^2(\mathbf{r}) m(\mathbf{r})\Big]+ \frac{1}{10}N \int \mathrm{d} \mathbf{r} \bigg[g_{\chi}|\chi(\mathbf{r})|^4 + \chi(\mathbf{r}) h_\varphi^*(\mathbf{r}) + \chi^*(\mathbf{r}) h_\varphi(\mathbf{r}) \bigg].
\end{align}
To simplify the notation, we define
\begin{equation}\label{def1}
     \eta(\mathbf{r}) = \frac{20}{9}gN |\chi(\mathbf{r})|^2, \quad g_\varphi = \frac{5}{9}gN, \quad h_{\chi}(\mathbf{r}) = \frac{10}{9}gN \chi(\mathbf{r})|\chi(\mathbf{r})|^2, 
\end{equation}
\noindent
\begin{equation}\label{def2}
    m(\mathbf{r}) = \frac{5}{9}gN\chi^2(\mathbf{r}), \quad
    g_{\chi} = 5gN, \quad
    h_{\varphi}(\mathbf{r}) =10gN\varphi(\mathbf{r}) |\varphi(\mathbf{r})|^2.
\end{equation}

Using Eqs.~(\ref{eq10}), (\ref{def1}), and (\ref{def2}) in Eq.~(\ref{eq8}), we obtain the expected value of the Hamiltonian of the Bose-Einstein condensate, denoted by $\mathcal{H}$. It is important to note that we are considering only the terms dependent on the condensate order parameter, $\varphi(\mathbf{r})$. The expected value $\mathcal{H}$ accounts solely for the contribution of the condensate’s kinetic energy, along with the energies associated with the trapping potential, the contact interaction between condensate atoms (self-interaction), and the interaction with the non-condensed gas cloud at the condensate’s surface. Therefore, we have:

\begin{align}\label{eq11}
    \mathcal{H} = \frac{9}{10}N \int \mathrm{d} \mathbf{r} &\left[\varphi^*(\mathbf{r}) \left(-\frac{\Delta}{2m} + V_{\mathrm{trap}}(\mathbf{r}) + \eta(\mathbf{r}) \right)\varphi(\mathbf{r}) + \varphi^2(\mathbf{r}) m^*(\mathbf{r})  \right. \nonumber \\
    &+ {\varphi^*}^2(\mathbf{r}) m(\mathbf{r})+ g_{\varphi}|\varphi(\mathbf{r})|^4+\varphi(\mathbf{r}) h_{\chi}^*(\mathbf{r})+ \varphi^*(\mathbf{r}) h_{\chi}(\mathbf{r}) \bigg],
\end{align}

 \noindent where $\eta(\mathbf{r})$, $h_{\chi}(\mathbf{r})$, and $m(\mathbf{r})$ are disorder fields. The Gross-Pitaevskii action functional is $S(\varphi, \varphi^*, h, h^*, m, m^*, \eta ) = S_0(\varphi, \varphi^*) + S_1(h, h^*, m, m^*, \eta)$. For simplicity, we denote $S_1(h, h^*, m, m^*, \eta) \equiv S_1$, we get:
\begin{align}\label{eq.12}
   S_0(\varphi, \varphi^*) + S_1 = \int d \mathbf{r} &\left[\varphi^*(\mathbf{r})\left(-\frac{\Delta}{2m} + m_0^2(\mathbf{r}) + \eta(\mathbf{r})\right)\varphi(\mathbf{r}) + g_\varphi |\varphi(\mathbf{r})|^4  \right.\nonumber\\ 
  & +\varphi(\mathbf{r})h^*(\mathbf{r}) +  \varphi^*(\mathbf{r})h(\mathbf{r})+  \varphi^2(\mathbf{r}) m^*(\mathbf{r}) + {\varphi^*}^2(\mathbf{r})m(\mathbf{r})\bigg],
\end{align}

\noindent where we redefined $h_{\chi}(\mathbf{r}) \equiv h(\mathbf{r})$. We also defined $m_0^2(\mathbf{r}) \equiv V_{\mathrm{trap}}(\mathbf{r}) - \mu_\varphi$ as the effective mass of a condensate particle. Additionally, it is possible to construct the Gross-Pitaevskii equation for the condensate, including both the linear and quadratic response terms, yielding the same result as described in Ref.~\refcite{Zaremba}. In the next section, we review the distributional zeta function method for the case of additive disorder.

\section{The distributional zeta function method}\label{sec:DZF}

From the action in Eq.~(\ref{eq.12}), we can define the disordered partition functional
\begin{equation}\label{ZZFM}
Z(h, h^*, m, m^*, \eta) = \int [d\varphi][d\varphi^*] e^{-S(\varphi, \varphi^*, h, h^*, m, m^*, \eta)}
\end{equation}
\noindent for the condensate. In Eq.~(\ref{eq.12}), $h(\mathbf{r})$, $m(\mathbf{r})$, and $\eta(\mathbf{r})$ may be interpreted as additive and multiplicative disorders, respectively. For completeness, in this section we present the distributional zeta-function method as
an approach to compute the average of $\ln Z(h,h^*)$. The results obtained here can be extended to the case of multiplicative disorder \cite{Acosta-Diaz:2019ntl}. By disregarding the multiplicative disorders in Eq.~(\ref{eq.12}) and substituting it into Eq.~(\ref{ZZFM}), the partition functional $Z(h,h^*)$ is written as:
\begin{align}\label{grossZ}
    Z(h, h^*) = \int [d\varphi][d\varphi^*] \exp&\left\{\int d \mathbf{r} \left[\varphi^*(\mathbf{r})\left(-\frac{\Delta}{2m} + m_0^2(\mathbf{r}) \right)\varphi(\mathbf{r}) + g_\varphi |\varphi(\mathbf{r})|^4 \right. \right. \nonumber\\ 
  &+\varphi(\mathbf{r})h^*(\mathbf{r})+  \varphi^*(\mathbf{r})h(\mathbf{r})\Bigg] \Bigg\}.
\end{align}

From the disordered functional $W(h, h^*) = \ln Z(h, h^*)$, we define the quenched free energy $\mathbb{E}[W(h, h^*)]$. Therefore,
\begin{equation}\label{QFE1}
    \mathbb{E}[W(h,h^*)] = \int [dh][dh^*] P(h,h^*)\ln Z(h, h^*),
\end{equation}

\noindent where $[dh][dh^*]P(h,h^*)$ is the probability distribution of the disorder. To proceed, let us discuss the distributional zeta-function $\Phi(s)$, defined by
\begin{equation}\label{phis}
    \Phi(s) = \int [dh][dh^*] P(h, h^*) \frac{1}{Z^s(h,h^*)}, \quad \mathrm{Re}(s)> 0.
\end{equation}

The description of the Bose-Einstein condensate is considered such that its action is invariant under the transformation $\varphi \rightarrow -\varphi$. Immediately, we have $Z(h, h^*)=Z(-h, -h^*)$, and one can see that $Z(0) \leq Z(h, h^*)$. Then, the function $\Phi(s)$ converges and is well-defined for the positive real part of the complex plane, and an analytic continuation is unnecessary. By comparing Eq.~(\ref{QFE1}) and Eq.~(\ref{phis}), we have that 
\begin{equation}\label{eqWds}
    \mathbb{E}[W(h,h^*)] = \left.-\frac{d}{ds}\Phi(s)\right|_{s\rightarrow0^+}.
\end{equation}

The average free energy can be represented by the following series of the
moments of the partition function:
\begin{equation}
    \mathbb{E}[W(h,h^*)] =  \sum_{k=1}^\infty \frac{(-1)^{k+1} c^k}{k! k} \mathbb{E}[Z^k(h, h^*)] + \log{c} + \gamma - R(c), \quad c>0\label{qfe}
\end{equation}

\noindent where $\gamma$ is the Euler–Mascheroni constant, and
    $\left| R(c) \right| \leq \frac{1}{Z(0)}\frac{e^{-Z(0)c}}{c}$. The contribution of
$R(c)$ to the free energy can be made as small as desired by taking $c$
large enough. In the next section, we will apply the distributional zeta-function method to the Bose-Einstein condensate with multiplicative disorder.

\section{Multiplicative disorder in the Bose-Einstein condensate}\label{sec:randommass}

In order to evaluate the quenched free energy with multiplicative disorder, it is useful to discuss some standard approximations to the complete action in Eq.~(\ref{eq.12}). Let us resort to approximations already made in the literature on the dynamics of the Bose-Einstein condensate, considering the coupling of its order parameter with the non-condensed degrees of freedom $\chi(\mathbf{r})$. Studies about the dynamics of the complete system usually resort to approximations that depend on the desired modeling. In Ref.~\refcite{GriffinPRB}, Griffin resorts to the well-known Hartree-Fock-Bogoliubov approximation by disregarding the three-field correlation function of the non-condensate gas in the Gross-Pitaevskii equation \cite{Bogolyubov:1959}. The three-field correlation function of the non-condensate is given by $h(\mathbf{r})$. In Ref.~\refcite{Hutchinson}, Hutchinson and Zaremba performed a Hartree-Fock-Bogoliubov-Popov approximation and disregarded the terms $m(\mathbf{r})$ and $h(\mathbf{r})$ in the Gross-Pitaevskii equation, keeping only the atomic density of the non-condensate, represented by $\eta(\mathbf{r})$. More about the Hartree-Fock-Bogoliubov-Popov approximation is presented in Ref.~\refcite{popov1987functional}. Here, we intend to consider Hartree-Fock-Bogoliubov and Hartree-Fock-Bogoliubov-Popov approximations in the condensate action presented in Eq.~(\ref{eq.12}). Our focus here is to model the contribution of $\eta(\mathbf{r})$ as a multiplicative disorder after disregarding $m(\mathbf{r})$ and $h(\mathbf{r})$ in the Gross-Pitaevskii functional for the condensate, by performing the Hartree-Fock-Bogoliubov-Popov approximation.

We start by discussing a Bose-Einstein condensate trapped in a three-dimensional volume $V$ by two parallel surfaces, which are separated by a distance $L$ along the $z$ axis, as shown in the illustrative diagram in Fig.~1. We might assume the presence of a non-condensed atomic gas cloud spreading into the region beyond the volume occupied by the Bose-Einstein condensate, which is bounded by the trapping potential. As previously mentioned, the amount of non-condensed gas accounts for only 10 percent of the total system. Given that the volume occupied by the non-condensed gas is much larger than that of the Bose-Einstein condensate, we can assume that most of the non-condensed gas resides outside the condensate. Consequently, the overlapping region has a smaller fraction of the non-condensate ($\ll 10\%$ of the total system); for this reason, it will be neglected in this work. We interpret the gas cloud as generating a surface disorder in the Bose-Einstein condensate, due to the interaction via the Hartree-Fock pseudo-potential between the condensate and the gas cloud.

\begin{figure}
    \begin{center}
    \includegraphics[width=0.9 \linewidth]{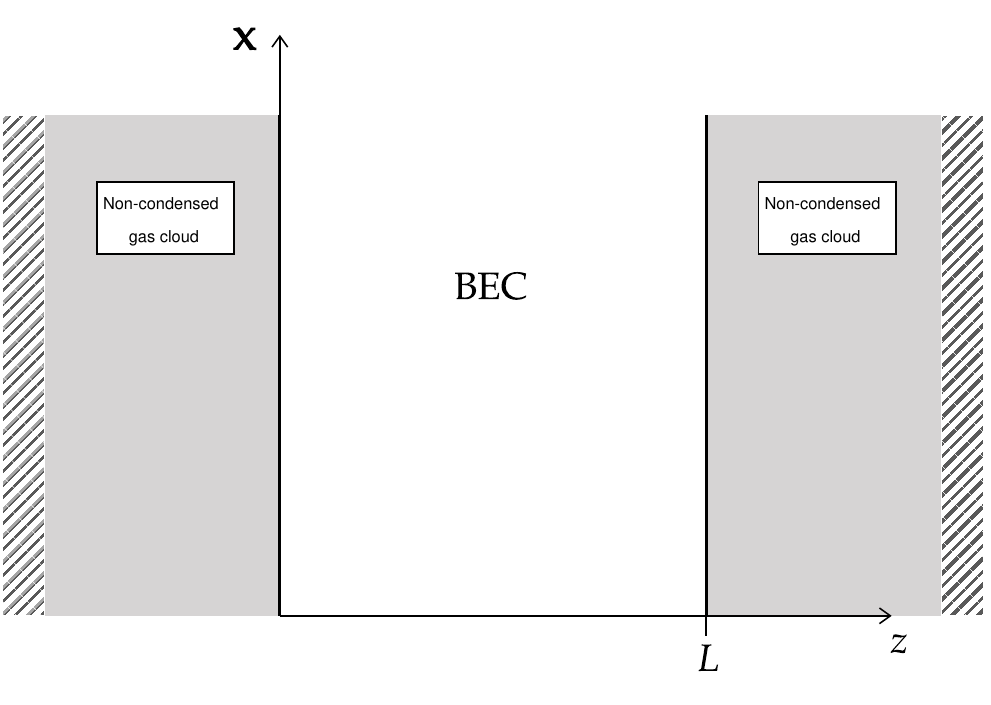}
    \end{center}
    \caption{Illustrative diagram of a Bose-Einstein condensate (BEC) with planar surfaces trapped in a region of length $L$. The non-condensed gas cloud is assumed to be located in the external region of the condensate's planar surfaces.}
    \label{Fig1}
\end{figure}

Now we use $\mathbf{r}=(\mathbf{x},z)$, where $\mathbf{x}$ represents the two-dimensional region where the planar surfaces of the condensate are defined. By considering the Hartree-Fock-Bogoliubov-Popov approximation, the Gross-Pitaevskii action for the Bose-Einstein condensate with the multiplicative disorder is
\begin{align}\label{eq12}
  S\left(\varphi, \varphi^*, \eta\right) = \int d^2\mathbf{x} \int^L_0 dz&\left\{\varphi^*(\mathbf{x},z)\left[-\frac{\Delta}{2m} + m_0^2(\mathbf{x},z) + \eta(\mathbf{x},z)\right]\varphi(\mathbf{x},z) \right.\nonumber\\
  &+ g_\varphi |\varphi(\mathbf{x},z)|^4\bigg\}.
\end{align}

We define the disordered functional $\displaystyle W(\eta)=\ln{Z(\eta)}$, just as presented in Sec.~\ref{sec:DZF}, in order to resort to the distributional zeta-function method. By considering the probability density of the disorder $P(\eta)$ defined as: 
\begin{equation}
    P(\eta) = p_0 \exp\left(-\frac{1}{2\rho^2}\int d^2\mathbf{x} \int_0^L dz \int  d^2\mathbf{x}' \int_0^L dz'\eta(\mathbf{x},z)F^{-1}(\mathbf{x},z;\mathbf{x}',z')\eta(\mathbf{x}',z')\right),
\end{equation}

\noindent where $\rho$ is the strength of the multiplicative disorder $\eta(\mathbf{x},z)$, and $F(\mathbf{x},z;\mathbf{x}',z')$ is the correlation function of the disorder. Therefore, the correlation of the disorder $\eta(\mathbf{x},z)$ is $\mathbb{E}[\eta(\mathbf{x},z)\eta(\mathbf{x}',z')] = \rho^2F(\mathbf{x},z;\mathbf{x}',z')$. According to Eq.~(\ref{qfe}), the effective action for the Bose-Einstein condensate is given by
\begin{align}\label{eq41}
    &S_{\textrm{eff}}(\varphi,\varphi^*) = \int d^2\mathbf{x} \,\int_0^Ldz \, \sum_{i=1}^k \varphi^*_i(\mathbf{x},z)\Bigg[-\frac{\Delta}{2m} + m_0^2(\mathbf{x},z)\Bigg]\varphi_i(\mathbf{x},z)  \nonumber\\
     &+  \int d^2\mathbf{x} \, d^2\mathbf{x}' \,\int_0^Ldz\int_0^L dz' \sum_{i,j=1}^k \, \left(\delta_{ij}g_\varphi-\rho^2 F(\mathbf{x},z;\mathbf{x}',z')\right) {\varphi^*_j}^2(\mathbf{x},z){\varphi_i}^2(\mathbf{x}',z').
\end{align}

The effective action $S_{\textrm{eff}}$ in Eq.~(\ref{eq41}) refers to the contribution of quadratic disorder in the condensate and is just a general result, since we have not yet defined exactly $F(\mathbf{x},z;\mathbf{x}',z')$. The random field $\eta(\mathbf{x},z)$ interacts with the Bose-Einstein condensate only on the planar surfaces. We model the interaction between the condensate and the non-condensed gas cloud as a Gaussian disorder distributed around the interface between the BEC and the non-condensate cloud. We can compute the result for $F(\mathbf{x},z;\mathbf{x}',z') = \delta^{2}(\mathbf{x}-\mathbf{x}')\left[\delta(z) + \delta(z-L)\right]$, since the planar surfaces of the condensate are located at $z=0$ and $z=L$. For this scenario, the $\delta^{2}(\mathbf{x}-\mathbf{x}')$ indicates a Gaussian distribution of the non-condensate along the surfaces of contact with the condensate. By considering the boundary condition $\varphi(\mathbf{x},0)=\varphi(\mathbf{x},L)$, we obtain:
\begin{align}\label{eq42}
    S_{\textrm{eff}}(\varphi,\varphi^*) 
     &=  \int d^2\mathbf{x} \,\int_0^L dz \, \sum_{i=1}^k \varphi^*_i(\mathbf{x},z)\Big[-\frac{\Delta}{2m} + m_0^2(\mathbf{x},z)\Big]\varphi_i(\mathbf{x},z) \nonumber\\
     & +  \int d^2\mathbf{x} \,\int_0^L dz \sum_{i,j=1}^k \Big(\delta_{ij} g_\varphi - 2 \rho^2 \Big) {\varphi_j}^{*2}(\mathbf{x},0){\varphi_i}^2(\mathbf{x},z).
\end{align}
Note that we obtained a non-local effective action \cite{sachdev, Aharony_2018}. Such a kind of theory does not allow us to interpret it as Euclidean wormhole effects in condensed matter systems, since the effective action in Eq.~(\ref{eq42}) is related to the quartic terms of $\varphi$. In the next section we will show that the situation is quite different for the case of additive disorder. 

\section{Euclidean wormholes in disordered condensate}\label{sec:randomfield}

The aim of this section is to discuss the situation in which the term $h(\mathbf{r})$ is not ignored, in the configuration of planar surfaces. Taking into account only the additive disorder, the partition function is given by:
\begin{align}\label{eq20}
    Z(h, h^*) &= \int [d\varphi][d\varphi^*] \exp\left\{-\int d^{2}\mathbf{x} \int_0^L dz \bigg[\varphi^*(\mathbf{x},z)\Big(-\frac{\Delta}{2m} + m_0^2(\mathbf{x},z) \Big)\varphi(\mathbf{x},z) \right.\nonumber\\
    &   + g_\varphi|\varphi(\mathbf{x},z)|^4  + \varphi(\mathbf{x},z)h^*(\mathbf{x},z) + \varphi^*(\mathbf{x},z)h(\mathbf{x},z) \bigg] \Bigg\}, 
\end{align}
Here we define the disorder correlation as $\mathbb{E}[h^*(\mathbf{x},z)h(\mathbf{x}',z')] = \sigma^2 F(\mathbf{x},z;\mathbf{x}',z')$, where $\sigma$ is the strength of the additive disorder $h(\mathbf{x},z)$. In our analyses, the non-Gaussian contribution is irrelevant. From Eq.~(\ref{qfe}), the Gaussian effective Gross-Pitaevskii action is:
\begin{align}\label{eq28}
    S_{\textrm{eff}}(\varphi,\varphi^*) &=  \int d^2\mathbf{x} d^2\mathbf{x}' \int_0^L dz \int_0^L dz' \sum_{i,j=1}^k  \varphi^{*}_j(\mathbf{x},z)\left[\left(-\frac{\Delta}{2m} + m_0^2(\mathbf{x},z)\right)\delta_{ij}\right.\nonumber\\
    &  \times \delta(\mathbf{x}-\mathbf{x}') \delta(z-z')  - \sigma^2 F(\mathbf{x},z;\mathbf{x}',z') \bigg]\varphi_i(\mathbf{x}',z').
\end{align}
Using the compact notation, where $\mathbf{r}=(\mathbf{x},z)$ and  $\mathbf{r}'=(\mathbf{x}',z')$, the operator
\begin{equation}
\mathcal{G}_{ij}(\mathbf{r},\mathbf{r}') = [-\Delta + m_0^2(\mathbf{x},z)] \delta_{ij} \delta^{3}(\mathbf{r}-\mathbf{r}') - \sigma^2 F(\mathbf{r},\mathbf{r}')
\end{equation}
is the $ij$-component of a $k\times k$ real symmetric matrix. The procedure for diagonalizing the effective action was addressed in more detail in Ref.~\refcite{Heymans:2024dzq}. Hence, the resulting diagonal matrix $\mathcal{G}_D(\mathbf{r},\mathbf{r}')$ is:

\begin{eqnarray}
    &\mathcal{G}_D(\mathbf{r},\mathbf{r'}) =
 \left(\begin{array}{cccc}
        \mathcal{G}_0(\mathbf{r},\mathbf{r'}) & 0 & \cdots & 0  \\
         0 &  \mathcal{G}_0(\mathbf{r},\mathbf{r'}) & \cdots & 0 \\
         \vdots & & \ddots & \vdots \\
         0 & \cdots & 0 & \mathcal{G}_0(\mathbf{r},\mathbf{r'}) - k \sigma^2 F(\mathbf{r},\mathbf{r'})
    \end{array}\right)_{k\times k}&
\end{eqnarray}

\noindent where $\mathcal{G}_0(\mathbf{r},\mathbf{r'}) = [-\Delta + m_0^2(\mathbf{x},z)]\delta^{3}(\mathbf{r}-\mathbf{r'})$. The structure of the diagonal matrix $\mathcal{G}_D(\mathbf{r},\mathbf{r'})$ allows us to separate the effective action into two contributions. One of the parts is related to the bare contribution to the connected two-point correlation functions even in the absence of disorder averaging, and was chosen to be related to the diagonal fields $\phi_1, ...,\phi_{k-1}$. The other part contains the contribution of $F(\mathbf{x},z;\mathbf{x}',z')$ acting on the $k$-th vector field. The splitting is possible because the modulus of the Jacobian is unity, since we have an orthogonal transformation. That is: 
\begin{equation}\label{ProdTransf}
    \prod_{a=1}^k [d\phi] \longrightarrow \prod_{a=1}^{k-1} [d\phi_a] [d\phi] \, \quad \mathrm{and} \quad \quad \prod_{a=1}^k [d\phi_a^{*}] \longrightarrow \prod_{a=1}^{k-1} [d \phi_a^{*}] [d\phi^*], 
\end{equation}

\noindent where $\phi$ in the equation above corresponds to the $k$-th field $\phi_k$. Therefore, one can write $\mathbb{E}[Z^k(h, h^*)]$ as:
\begin{equation}
    \mathbb{E}[Z^k(h, h^*)] = \int \prod_{i=1}^{k-1} [d\phi_a][d\phi^{*}_a]\,e^{-S_0\left(\phi_a, \, \phi_a^{*}\right)} \int [d\phi][d\phi^*] e^{-S_\sigma \left(\phi, \, \phi^* \right)},
\end{equation}

\noindent where $S_0\left(\phi_a, \, \phi_a^{*}\right)$ is the total action for the fields $\phi_1, ..., \phi_{k-1}$ and is given by
\begin{equation}\label{eq:acfree}
     S_0\left(\phi_a, \, \phi_a^{*}\right) = \int d^2\mathbf{x} \, \int_0^L dz \, \, \sum_{a=1}^{k-1} \phi_a^{*}(\mathbf{x},z)\left[-\Delta + m_0^2(\mathbf{x},z)\right]\phi_a(\mathbf{x},z) 
\end{equation}

\noindent and $S_\sigma \left(\phi, \, \phi^{*}\right)$ is the action for the $k$-th field that encodes the disorder correlation function $F(\mathbf{x},z;\mathbf{x}',z')$. The functional $S_\sigma \left(\phi, \, \phi^{*}\right)$ is
\begin{eqnarray} \label{sigmaAction}
    S_\sigma \left(\phi,  \phi^{*}\right) &=&  \int d^2\mathbf{x} d^2\mathbf{x}'\int_0^L dz \int_0^L dz' \phi^*(\mathbf{x},z)\Big\{\left[-\Delta+m_0^2(\mathbf{x},z)\right]\delta(\mathbf{x}-\mathbf{x}')\delta(z-z') \nonumber\\
     &&- k\sigma^2 F(\mathbf{x},z;\mathbf{x}',z')\Big\}\phi(\mathbf{x}',z').
\end{eqnarray}

Note that we have not yet defined $F(\mathbf{x}, z; \mathbf{x}',z')$ in Eq. (\ref{sigmaAction}). First, we can resort to the definition for $F(\mathbf{x},z;\mathbf{x}',z')$ similar to the way mentioned in Sec. \ref{sec:DZF}. Now we choose the covariance as $F(\mathbf{x},z;\mathbf{x}',z')= \delta^{2}(\mathbf{x}-\mathbf{x}')\left[\delta(z) + \delta(z-L)\right]$. With this choice, the action becomes
\begin{equation}\label{eq31}
    S_{\sigma}\left(\phi, \, \phi^* \right) = S_{G}\left(\phi, \, \phi^* \right) -k\sigma^2\int d^2\mathbf{x}\int_0^L dz  \, \phi(\mathbf{x},z)\big[\phi^*(\mathbf{x},0)+\phi^*(\mathbf{x},L)\big]
\end{equation}

\noindent where $S_{G}\left(\phi, \, \phi^* \right) $ refers to the integration with the operator $\left[-\Delta + m_0^2(\mathbf{x},z)\right]$ in Eq. (\ref{sigmaAction}). The second integral in Eq. (\ref{eq31}) is a series of connections between the points on the surfaces and other internal points of the condensate in the same direction in $\mathbf{x}$. In addition to the connections between the points on one of the surfaces and points in the internal region, there are connections between two points on the two parallel surfaces of the Bose-Einstein condensate. 

However, for an analog model of Euclidean wormholes, the contribution of $F(\mathbf{x},z;\mathbf{x}',z')$ is interesting when it is not $\delta$-correlated along the $z$ axis. Therefore, consider $F(\mathbf{x},z;\mathbf{x}',z') = \delta^{2}(\mathbf{x}-\mathbf{x}')C(z-z')$, in which $C(z-z')$ is expected to encode the non-locality. Therefore, Eq. (\ref{sigmaAction}) becomes:

\begin{equation}\label{eq30}
    S_{\sigma}\left(\phi, \, \phi^{*}\right) = S_G\left(\phi, \, \phi^* \right)  -k\sigma^2 \int d^2\mathbf{x} \int_0^L dz\int_0^L dz' \, \,  \phi^*(\mathbf{x},z)C(z-z')\phi(\mathbf{x},z')
\end{equation}

The second integral in Eq. (\ref{eq30}) indicates the non-local connections between two points on the $z$ axis in the condensate. Due to the Gaussian distribution of disorder along the $\mathbf{x}$ direction, the non-local connection will only be between points that are at the same $\mathbf{x}$ coordinate. The above result shows similarity to the action for Euclidean quantum gravity (see Eq. (\ref{eq.1})). In this comparison, $C(z-z')$ would play the same role as $C_{ij}(x,y)$ in Eq. (\ref{eq.1}), thus configuring the Bose-Einstein condensate interacting with the non-condensed atomic gas cloud as an analogous model for Euclidean wormholes. One can say that the disorder average of the free energy leads to a superposition of contributions of many points of condensate connected by Euclidean wormholes.

\begin{figure}
    \centering
    \includegraphics[width=0.8\linewidth]{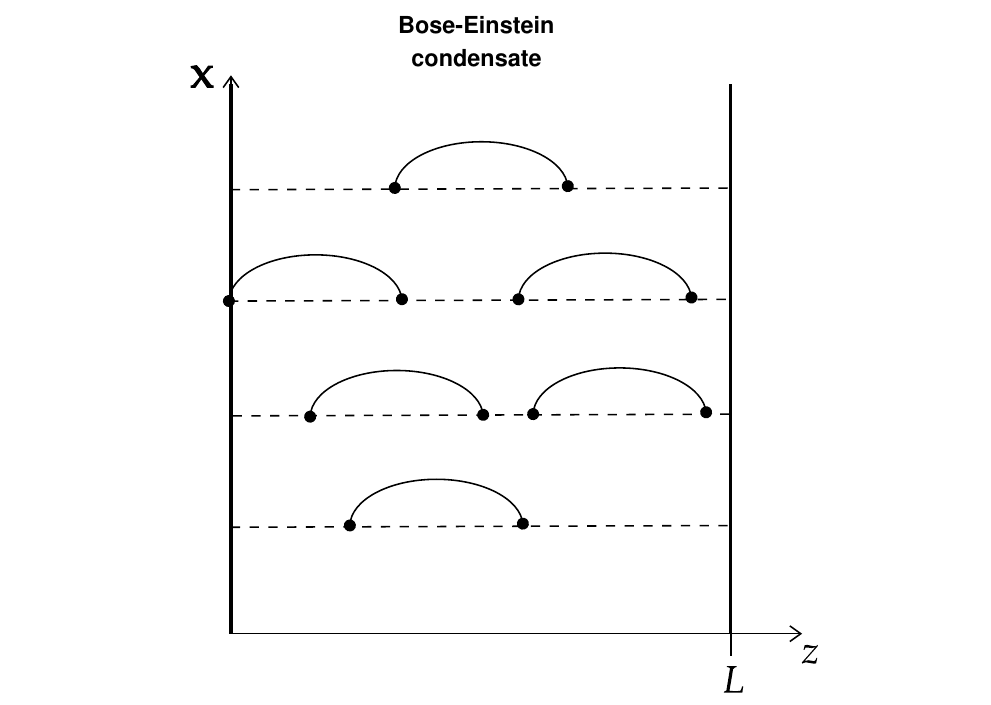}
    \caption{Illustrative diagram of non-local connections in the disordered Bose-Einstein condensate. These non-local connections are described by the same mathematical structure as Euclidean wormholes.}
    \label{fig2}
\end{figure}
 
\section{Casimir pressure in a disordered Bose-Einstein condensate}\label{Sec:Cas}

It is well known that quantum fluctuations of the electromagnetic field may generate an attractive or repulsive interaction between two closely spaced electrically neutral surfaces \cite{Casimir:1948dh}, known as the Casimir force. Such an effect was first observed in parallel plates geometry \cite{SKLamo, Bressi_2002}. Furthermore, in a perfect Bose gas confined to a slab geometry, thermal fluctuations give rise to a pressure associated with a force analogous to the Casimir effect \cite{PAMartin}. Considering periodic boundary conditions in a weakly-interacting Bose-Einstein condensate within parallel plates geometry, Ref. [\refcite{Edery_2006}] obtains the following Casimir pressure  
\begin{equation}\label{press}
    P \propto -\frac{7\pi^2}{180 d^4},
\end{equation}

\noindent where $d$ is the value of the plate separation and the negative sign means that the force related to this pressure is attractive. 

Here, using the construction of the previous sections, we calculate the Casimir pressure for the Bose-Einstein condensate in the presence of the non-condensed gas cloud, where the non-condensed cloud is modeled by a non-local additive quenched disorder. On more physical grounds, we can model these plates as a permeable structure to the non-condensed cloud but confining the BEC. Also, we can interpret an emergent force over the non-condensed gas as a Casimir effect due to the fluctuations in the condensate. In this sense, we are calculating the effect of the non-local interactions (which can be interpreted as the wormholes of our analog model) on the Casimir pressure in the Bose-Einstein condensate. Using the action of Eq. (\ref{eq30}), we can obtain the following equation of motion:
\begin{equation}\label{moteq}
    \left[-\Delta + m_0^2(\mathbf{x},z)\right]\phi(\mathbf{x},z) - k\sigma^2 \int_0^L dz'\, C(z-z')\phi(\mathbf{x},z') = 0.
\end{equation}

In order to proceed, we can consider the simplest form of $C(z-z')$:
\begin{equation}
    C(z-z') =\left\{\begin{array}{ccc}
        b_1 & &\mathrm{if} \,\,z>z', \\
        b_2 & &\mathrm{if} \,\,z'>z,
    \end{array}\right.
\end{equation}

\noindent where $b_1$ and $b_2$ are two different real constants. To solve the non-locality problem in Eq. (\ref{moteq}), we can resort to the definition of fractional derivative \cite{Diethelm2010}: 
\begin{equation}
    D_\mu^\nu \phi(x,z) = \frac{1}{\Gamma(\nu)}\int_\mu^z dz'\, \phi(\mathbf{x},z')(z-z')^{\nu-1}, \quad 0<\nu\leq 1.
\end{equation}
Therefore, Eq. (\ref{moteq}) becomes:
\begin{equation}
    \left[-\Delta + m_0^2(\mathbf{x},z) + k\sigma^2\left(b_1D_0^1 - b_2D_L^1\right)\right]\phi(\mathbf{x},z) = 0.
\end{equation}

We have defined before $m_0^2(\mathbf{x},z) = V_{trap}(\mathbf{x},z) - \mu$. If we consider a constant value for $V_{trap}$, then $m^2_0(\mathbf{x},z)$ is also a positive definite constant value $m^2_0$. Let us suppose that $b_2>b_1$. Proceeding in an analogous way to Ref. [\refcite{Heymansrestoration}], we can obtain the Fourier representation of the two-point correlation function of the $k$-th equation of motion, Eq. (\ref{moteq}), as:
\begin{equation}\label{funcGren}
    G_0^{(k)}(\mathbf{q}_\mathbf{x}, q_z) = \frac{1}{\mathbf{q}_\mathbf{x}^2 + q_z^2 + k\sigma^2\epsilon|q_z|+ m_0^2},
\end{equation}

\noindent in which $\epsilon\equiv |b_2-b_1|$ is a positive value. From Eq. (\ref{funcGren}), we obtain the spectrum associated with the energy of the disordered condensate. The Casimir energy can be obtained through the derivation of the analytic regularization of the spectral zeta function, $\zeta_\mathcal{O}(s)$, constructed from the energy spectrum of the Bose-Einstein condensate, i.e.,
\begin{equation}\label{zeta1}
    \zeta_\mathcal{O}(s) = \sum_{\mathbf{q}_\mathbf{x}, q_z} \left(\mathbf{q}_\mathbf{x}^2 + q_z^2 + k\sigma^2\epsilon|q_z|+ m_0^2\right)^{-s}.
\end{equation}
From the last equation, it is immediate to see that the Casimir energy is related to $\zeta_\mathcal{O}(s = -1/2)$ \cite{Blau:1988kv}.
We consider Dirichlet boundary conditions at the surfaces of the Bose-Einstein condensate along the $z$-direction, so the momentum $q_z$ assumes discrete values, $q_z = \frac{n\pi}{L}$. The integral form of the expression for $\zeta_\mathcal{O}(s)$ is
\begin{eqnarray}\label{zetA2}
    \zeta_\mathcal{O}(s) &=& \frac{A}{2\pi} \sum_{n=1}^\infty \int dq \, q \left[q^2 + \left(\frac{n\pi}{L}\right)^2 + k\sigma^2 \epsilon \frac{n\pi}{L} + m_0^2\right]^{-s} \nonumber\\
    &=& \frac{A}{2\pi}\frac{L^{2s}}{\pi^{2s}}\sum_{n=1}^{\infty}\int_0^\infty dq \, q \left[\frac{L^2 q^2}{\pi^2} + n^2 +  \frac{k\sigma^2 \epsilon L}{\pi}n + \frac{m_0^2L^2}{\pi^2}\right]^{-s},
\end{eqnarray}

\noindent where $A$ is the area of the plates and $q\equiv |\mathbf{q}_\mathbf{x}|$. Using the following Mellin representation:
\begin{equation}
    a^{-s} = \frac{1}{\Gamma(s)}\int_0^\infty dx\, x^{s-1} e^{-ax},
\end{equation}
we can recast Eq. (\ref{zetA2}) as
\begin{equation}\label{eq:zetas}
    \zeta_\mathcal{O}(s) = \frac{AL^{2s-2}}{4\pi^{2s-1}\Gamma(s)}\int_0^\infty dx\, x^{s-2}  \sum_{n=1}^{\infty} \exp{\left[-x\left(n^2 + \frac{k\sigma^2\epsilon L}{\pi}n + \frac{m_0^2 L^2}{\pi^2}\right)\right]}.
\end{equation}
Note that the previous equation is similar to Eq. (55) of Ref. [\refcite{Heymansrestoration}]. Therefore, we can proceed analogously to regularize Eq. (\ref{eq:zetas}). As explicitly obtained in Ref. [\refcite{Heymansrestoration}], only a set of $k$'s can be regularized; such a set is given by $k=  \left\lfloor \frac{2m_0}{\sigma^2 \epsilon}\right\rfloor l,\,\, l \in \mathbb{N}$, where $\lfloor \alpha \rfloor$ denotes the largest integer $\leq \alpha$. The Casimir effect arises when the correlation length is at least of the same order of magnitude as the distance between the plates. Therefore, the main contribution to the Casimir energy will arise for large correlation lengths. In our case, such a scenario is obtained when we fix $l = \left\lfloor\frac{ \sqrt{m^2_0L^2}}{\pi}\right\rfloor = l_0$; therefore, the main contribution to the Casimir energy will be related to
\begin{equation}\label{zetafinal}
     \zeta_\mathcal{O} (s)= \frac{AL^{2s-2}}{4\pi^{2s-1}}\frac{\Gamma(s-1)}{\Gamma(s)}\sum_{n=1}^\infty \left(n+ l_0\right)^{2-2s}, \quad l_0 \in \mathbb{N}.
\end{equation}

Using that $\Gamma(s) = (s-1)\Gamma(s-1)$ in the above equation, the summation in Eq. (\ref{zetafinal}) is related to the Hurwitz zeta function $\zeta_H\left(2s-2, l_0\right)$.

The spectral zeta function at $s=-1/2$ is given by
\begin{equation}\label{CasEnergy}
    \zeta_\mathcal{O} \left(-\frac{1}{2}\right) = -\frac{A}{6L^3}\left[\zeta_H(-3,l_0) - l_0^3 \right].
\end{equation}
 
To be consistent, the Hurwitz zeta function needs to be taken as its analytic continuation \cite{elizalde}, which is 
\begin{equation}
    \zeta_H (1-m, y) = -\frac{B_m(y)}{m},\quad m \in \mathbb{N},
\end{equation}
where $B_m(y)$ is the $m$-th Bernoulli polynomial. Therefore, it leads us to
\begin{equation}
    \zeta_\mathcal{O} \left(-\frac{1}{2}\right) = \frac{1}{24L^3}\left[l_0^2(l_0 + 1)^2 - \frac{1}{30} \right].
\end{equation}

Now, some care must be taken to define our final Casimir energy. First, we must remember that we do have the free energy of the system represented by a series, see Eq. (\ref{eqWds}). Therefore, we can proceed with the same reasoning as in Ref. [\refcite{Heymans:2024dzq}], and the Casimir energy will be given by 
\begin{equation}
    E_c = \frac{(-1)^{l_0}}{l_0 l_0!}\exp\left[l_0 \ln c - \left.\zeta_{\mathcal{O}}(s)\right|_{s=-\frac{1}{2}} \right],
\end{equation}
so we must choose $c$ to maximize the exponential. Note that in the previous equation, we neglected the contribution of the action in Eq. (\ref{eq:acfree}) because we are disregarding finite size effects; thus, such an action always has an exponentially decaying correlation function, which does not contribute to the Casimir effect. Defining the Casimir pressure as the Casimir force per area of the plates, we can write
\begin{equation}
    P_c = -\frac{1}{A}\frac{\partial E_c}{\partial L} = \frac{(-1)^{l_0+1}}{2l_0l_0!}\frac{\partial}{\partial L} \zeta_{\mathcal{O}}\left(s=-\frac{1}{2}\right).
\end{equation}

\begin{figure}
    \centering
    \includegraphics[width=0.7\linewidth]{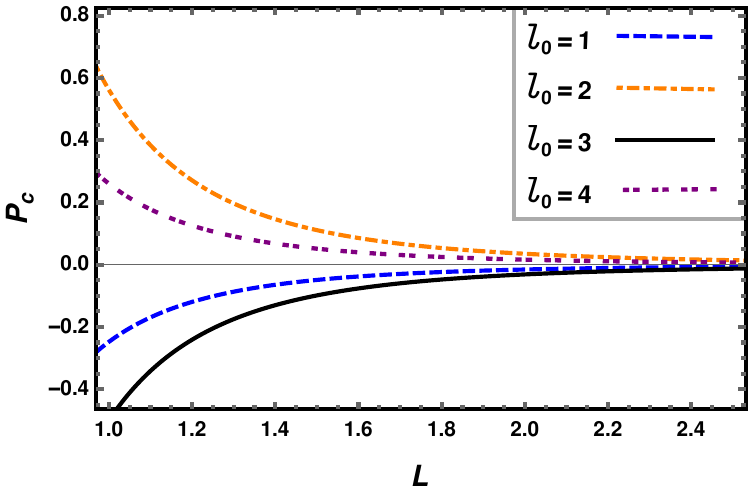}
    \caption{Plot of the Casimir pressure $P_c$, Eq.~(\ref{eq:CasPre}), for $l_0 = 1, 2, 3, 4$.}
    \label{fig3}
\end{figure}

From the previous equations, it follows directly that the leading contributions to the Casimir pressure, when one takes into account the effects of the non-condensed cloud, are given by
\begin{equation}\label{eq:CasPre}
    P_c = \frac{(-1)^{l_0}}{16 l_0 l_0!} \frac{1}{L^4} \left[l_0^2 (l_0 + 1)^2 - \frac{1}{30} \right].
\end{equation}
Thus, the sign of $P_c$ depends on which moments of the distributional zeta function maximize the correlation length, see Fig.~\ref{fig3}, as expected in light of Refs.~[\refcite{Rodriguez-Camargo:2022wyz,Heymans:2024dzq}]. In conclusion, we see that the effects of the non-condensed cloud (the analog counterpart of Euclidean wormholes) can modify the characteristic attractive force in the Casimir effect of the Bose-Einstein condensate depending on experimental parameters such as the system size. 

\section{Conclusions}\label{sec:conc}

Many approaches to quantum gravity discuss modifications of standard physics at short distances. However, it has become increasingly clear that short-distance modifications alone cannot address many fundamental problems in quantum gravity. One must understand long-distance physics and even topology change. Instead of discussing analog models for lightcone fluctuations in a pseudo-Riemannian manifold, we adopt the Euclidean viewpoint, which is a central tool in the mathematical study of quantum field theory. Many authors argue that quantum gravity must be formalized in a Euclidean domain. 

In a Euclidean theory of quantum gravity, it is expected that the topology of spacetime can vary, allowing the description of black holes and closed universes. Such a theory should accommodate topologies related to closed universes that branch off from or join flat spacetime. In this context, wormholes connect two asymptotically Euclidean regions or two parts of the same asymptotically Euclidean region. A non-local contribution to the effective action was obtained, thereby proposing an analog model for Euclidean wormholes. In the present work, we applied this scenario to a disordered Bose-Einstein condensate.

In Refs.~\refcite{zaremba1999,Zaremba}, the authors emphasized that the presence of a thermal cloud of non-condensed atoms may give rise to new phenomena. The crucial question is the coupling between the degrees of freedom of the condensate and the normal fluid. Here, we discuss the effect of the non-condensed atomic cloud on the Bose-Einstein condensate. First, we demonstrate that the resulting linear term from the interaction between the condensate and the atomic gas can be interpreted as quenched disorder. Using the distributional zeta-functional method, we obtained a representation for the quenched free energy as a series of integral moments of the partition function. Assuming two planar surfaces confining the Bose-Einstein condensate, we show that random surface fields generate non-local terms in the effective action. We argue that the interaction between the Bose-Einstein condensate and the surrounding atomic cloud can serve as an analog model for non-local effects in condensed matter systems, described by the same mathematical structure as Euclidean wormholes. 

Finally, to investigate how nonlocal connections—analogous to Euclidean wormholes—might influence physical quantities in a Bose-Einstein condensate, we proposed a calculation of the Casimir pressure via zeta function regularization, based on the energy spectrum of the system illustrated in Fig.~1. We have shown that a Casimir pressure $P_c$ arises in the Bose-Einstein condensate due to disorder, as obtained via the analytic continuation of the spectral zeta function. For simplicity, we calculate the Casimir pressure associated with a specific multiplet. This contribution can be either attractive or repulsive.

A natural continuation of this work is to study in detail the case of multiplicative disorder. This subject is currently under investigation by the authors.

\section{Acknowledgments}

The authors thanks to the referees, whose comments and
suggestions lead to a better presentation of our work. I. P. F. thanks M. M. Balbino, F. S. Sorage, and J. T. Miranda for useful discussions. G. O. H. thanks Fundação Carlos Chagas Filho de Amparo à Pesquisa do Estado do Rio de Janeiro (FAPERJ) and Coordenação de Aperfeiçoamento de Pessoal de Nível Superior (CAPES) for financial support. This work was partially supported by Conselho Nacional de Desenvolvimento Científico e Tecnológico (CNPq), grants nos. 305000/2023-3 (N.F.S). I. P. F. also thanks Conselho Nacional de Desenvolvimento Científico e Tecnológico (CNPq) for financial support.

\bibliographystyle{ws-ijmpd}

\end{document}